\documentclass[12pt]{article}

\usepackage{times}
\usepackage{geometry}
\usepackage{graphicx}
\usepackage[utf8]{inputenc}
\usepackage{enumitem,amssymb,amsmath}
\usepackage{ragged2e}
\usepackage{wrapfig}
\usepackage{color}
\newlist{thematic}{itemize}{8}
\setlist[thematic]{label=$\square$}
\usepackage{pifont}
\usepackage{url}
\usepackage[nottoc]{tocbibind}
\usepackage[numbers]{natbib}
\usepackage{xcolor}
\usepackage{multicol}
\definecolor{navy}{RGB}{0,0,130}
\usepackage[colorlinks=true,allcolors=navy,linktocpage]{hyperref}
\usepackage{mathptmx}
\usepackage[font=small,skip=0pt]{caption}

\usepackage{tocloft}

\newcommand{\detector}{AMIGO}
\renewcommand{\paragraph}[1]{\vspace{0.4em} \noindent{\bf #1}}

\begin{document}
{\RaggedRight
\fontsize{17.75}{20}\selectfont
A Voyage 2050 Science White Paper Submission \linebreak \linebreak
\fontsize{26.75}{32}\selectfont
\RaggedRight  \bf \em  Probing the Nature of Black Holes:
\\ Deep in the mHz Gravitational-Wave Sky
\vfill }


\normalsize
\noindent\textbf{Contact Scientist:} Vitor Cardoso \\
\noindent Primary institution and address: CENTRA, Instituto Superior Técnico, Universidade de Lisboa, Avenida Rovisco Pais 1, 1049 Lisboa, Portugal \\
Email: vitor.cardoso@tecnico.ulisboa.pt \\
Phone:  351-218419821 \\

\noindent\textbf{Abstract:}
 Black holes are unique
  among astrophysical sources: they are the simplest macroscopic  objects in the Universe, and they are extraordinary in terms of their ability to convert energy into electromagnetic and gravitational radiation. Our capacity to probe their nature is limited by the sensitivity of our detectors. The LIGO/Virgo interferometers are the gravitational-wave equivalent of Galileo's telescope. The first few detections represent the beginning of a long journey of exploration.
  At the current pace of technological progress, it is reasonable to expect that the gravitational-wave detectors available in the 2035-2050s will be formidable tools to explore these fascinating objects in the cosmos, and space-based detectors with peak sensitivities in the mHz band represent one class of such tools. These detectors have a staggering discovery potential, and they will address fundamental open questions in physics and astronomy. Are astrophysical black holes adequately described by general relativity?  Do we have empirical evidence for event horizons? Can black holes provide a glimpse into quantum gravity, or reveal a classical breakdown of Einstein's gravity? How and when did black holes form, and how do they grow? Are there new long-range interactions or fields in our universe, potentially related to dark matter and dark energy or a more fundamental description of gravitation? Precision tests of black hole spacetimes with mHz-band gravitational-wave detectors will probe general relativity and fundamental physics in previously inaccessible regimes, and allow us to address some of these fundamental issues in our current understanding of nature.
  
\thispagestyle{empty}
\pagebreak

\noindent\textbf{Members of the proposing team:} \\
Vishal	Baibhav	(Johns Hopkins U, Baltimore, Maryland, USA)	,\\
Leor        Barack	(U. Southampton, UK)	,\\
Emanuele	Berti	(Johns Hopkins U., Baltimore, Maryland, USA)	,\\
Béatrice	Bonga	(Perimeter Institute, Waterloo, CA)	,\\
Richard	Brito	(U. Rome ``La Sapienza'', Rome, IT)	,\\
Vitor	Cardoso$^\dagger$	(IST, U. Lisboa, PT)	,\\
Geoffrey	Compère	(U. Libre de Bruxelles, BE)	,\\
Saurya	Das	(U. Lethbridge, CA)	,\\
Daniela 	Doneva	(Eberhard
Karls U. T\"ubingen, DE)	,\\
Juan	Garcia-Bellido	(U. Autonoma Madrid, ES)	,\\
Lavinia	Heisenberg	(Institute  for  Theoretical  Physics,  ETH  Z\"urich, CH)	,\\
Scott A. Hughes	(Massachusetts Institute of Technology, Cambridge, USA)	,\\
Maximiliano	Isi	(Massachusetts Institute of Technology, Cambridge, USA)	,\\
Karan	Jani	(Georgia Institute of Technology, Atlanta, USA)	,\\
Chris	Kavanagh	(Max Planck Institute for Gravitational Physics, Potsdam, DE)	,\\
Georgios	Lukes-Gerakopoulos	(Astron. Inst., Czech Acad. Sci., Prague, CZ)	,\\
Guido Mueller (U. Florida, Gainesville, USA)	,\\
Paolo	Pani	(U. Rome ``La Sapienza'', IT)	,\\
Antoine Petiteau	(APC, CNRS / U. Paris-Diderot, FR)	,\\
Surjeet	Rajendran	(Johns  Hopkins U., Baltimore, Maryland, USA)	,\\
Thomas P.~Sotiriou	(U.~of Nottingham, UK)	,\\
Nikolaos	Stergioulas	(Aristotle U. of Thessaloniki, GR)	,\\
Alasdair	Taylor	(U. Glasgow, UK)	,\\
Elias	Vagenas	(Kuwait U., Safat, KW)	,\\
Maarten	van de Meent (Max Planck Institute for Gravitational Physics, Potsdam, DE)	,\\
Niels	Warburton	(U. College Dublin, IE)	,\\
Barry	Wardell	(U. College Dublin, IE)	,\\
Vojt{\v e}ch	Witzany$^\dagger$	(Astron. Inst., Czech Acad. Sci., Prague, CZ)	,\\
Aaron	Zimmerman	(U.  Texas  at  Austin,  USA) \\
\vfill
{\small \noindent
$^\dagger$ Principal coordinators

\thispagestyle{empty}
\pagebreak
\setcounter{page}{1}
\tableofcontents
}

\section{Introduction}\label{introduction}
%

Our understanding of nature has always been driven by observations and new data.  The Standard Model of particle physics was established through the long-term support for particle accelerators of higher and higher center-of-mass energies.  With the advent of direct gravitational-wave detection, our understanding of gravity (and possibly of other fundamental interactions) is now undergoing a phase transition~\cite{Barack:2018yly}. 

Experimentally, this progress was largely driven by the historical detection of gravitational waves (GWs) by the LIGO/Virgo collaboration~\cite{Abbott:2016blz}.  The
current network of GW interferometers is rapidly expanding on Earth (with the construction and operation of KAGRA~\cite{Akutsu:2017thy} and LIGO-India~\cite{Unnikrishnan:2013qwa}) and in space.
In 2013, ESA’s Science Programme Committee selected {\em The
  Gravitational Universe} (millihertz GW astronomy) as the science theme for the 3rd large-class mission in the Cosmic Vision Programme (L3). A call for mission concepts was issued by ESA in 2016. Following the spectacular success of LISA Pathfinder~\cite{Armano:2016bkm}, a
European-US team of scientists responded with a proposal in support of the Laser Interferometer Space Antenna (LISA)~\cite{Audley:2017drz}, which was selected by ESA in June 2017. LISA will be the pioneering mission unveiling the mHz gravitational-wave sky. 

In the meanwhile, X-ray
spectroscopy~\cite{Reynolds:2013qqa,Risaliti:2013cbe} and
continuum-fitting methods~\cite{McClintock:2011zq} provide access to regions within a few Schwarzschild radii of compact objects, enabling us to measure spins more accurately than ever before. Novel astronomical techniques such as radio and deep infrared
interferometry~\cite{Doeleman:2008qh,Antoniadis:2013pzd} are giving us direct {\it images} of the dark, massive and compact object lurking at the center of  our galaxy~\cite{Genzel:2010zy,2018A&A...618L..10G,Akiyama:2019cqa}. The exquisitely precise timing of binary
pulsars~\cite{Antoniadis:2013pzd,Porayko:2018sfa} and
NICER~\cite{2017NatAs...1..895G} are providing incredibly accurate measurements of gravity in the strong-field regions surrounding neutron stars (NSs). 
Not since the 1960s, when the observations of the first quasars and the birth of X-ray astronomy revolutionized our understanding of high-energy astrophysics, have we witnessed such an experimental renaissance of general relativity (GR). The wealth of
information and huge discovery potential offered by current and future GW observatories has had an impact on the global scientific community. Gravitational physics has become a data-driven research field, ready to spearhead the effort towards answering some of the most challenging open questions in 
science. Gravity has also become a melting pot for different communities -- including relativists, particle theorists, astronomers, cosmologists,  data scientists, and experimentalists -- with the common goal to extract new physics and astrophysics from
GW observations, and to understand which theories or predictions can
be tested with the new data.

Central to this new era in physics and astronomy are black holes (BHs), the most compact objects in the Universe, with masses ranging from a few to a few billion solar masses.
BHs are arguably the ideal physics laboratory. 
According to GR, BHs are the simplest macroscopic
objects, being fully characterized by only a small set of
parameters. Quoting Subrahmanyan Chandrasekhar~\cite{Chandra1}: {\em ``In my entire
  scientific life, extending over forty-five years, the most
  shattering experience has been the realization that an exact
  solution of Einstein's equations of general relativity provides the
  absolutely exact representation of untold numbers of black holes
  that populate the universe.''}
BH physics is therefore ``clean,'' and less prone to systematic or astrophysical uncertainties.
Some BH binaries will slowly inspiral emitting GWs in the millihertz band for years, mapping the spacetime geometry to exquisite precision. Other BH binaries will
merge and emit, in this same band, loud ``ringdown'' radiation whose properties characterize the structure of the final BH (``BH spectroscopy''). Resolving both signals will allow us to test the predictions of GR and to look for new physics. Electromagnetic counterparts will inform us on the environment that these BHs live in, such as the properties of their accretion disks. Any deviation that cannot be accounted for by systematics would signal the presence of exciting new physics that would need to be modelled in some beyond-GR scenario in order to be properly accounted for and interpreted.

Simple as BHs may be, they have some remarkable features, such as horizons and ergoregions, that lend further support to the idea of using them as natural laboratories. Horizons are membranes shielding us from regions where classical GR breaks down, and are associated with long-standing paradoxes in theoretical physics,
such as the potential loss of unitarity due to thermal Hawking radiation. Certain proposals for resolving these paradoxes rely on the assumption that new physics appears at horizon scales, motivating further observational attempts to probe BH horizons. 
Particles or waves that enter the ergoregion of an astrophysical, rotating BH can escape to infinity with more energy than they had when they fell in, mining angular momentum from the BH. The associated amplification effects are of great interest both astrophysically (e.g., in the context of relativistic jets) and from a fundamental physics perspective, because they could lead to the experimental observation of light bosonic fields. 

In this White Paper we will focus on the fundamental physics science case of a more sensitive detector operating in the millihertz band that we will call the Advanced Millihertz Gravitational-wave Observatory, or {\em \detector{}} for short. Such a detector would observe large numbers of BH systems with total mass in the range $10^4 M_\odot < M< 10^9 M_\odot$, shedding light on the population of supermassive BHs (SMBHs) residing in the centers of galaxies and on the intermediate-mass BHs (IMBHs) which may exist in the centers of some sub-galactic structures and clusters. SMBHs are major cosmic players: they are tiny compared to galactic scales, but the associated jets and accretion disks can often outshine the entire host galaxy. The properties of SMBHs are tightly correlated with those of their host galaxies, and therefore they are believed to play a central role in the dynamics of stars and gas at galactic scales. However, the details of the birth and growth of SMBHs are not well understood. 
There is even more uncertainty at the low-mass end, where a conclusive detection of IMBHs has not yet been made, and their number and properties are therefore unknown. A sensitive millihertz GW detector such as \detector{} will also inform us on the environments surrounding these BHs, and give us a census of their masses and spins throughout cosmic history.

About 400 years after Galileo's first exploration of telescopes, we are just beginning to understand GW technology
and the information it can provide. To interpret the wealth of data from GW observations and experiments in the next decades we need a comprehensive theoretical effort in the numerical and analytical modeling of GW sources. Together with ongoing efforts on the theoretical side, powerful new GW detectors truly realize the immense potential of GW physics. The science case for a sensitive GW-detector in the millihertz band is cross-cutting and multidisciplinary, and will address outstanding open issues in physics:
\begin{description}
\item[$\cdot$ Is GR the correct description for strong and dynamical gravitational interactions?] 

\item[$\cdot$ Are there fundamental fields beyond the Standard Model of particle physics?] 

\item[$\cdot$ What are the distribution and properties of dark matter close to galactic centers?]

\item[$\cdot$ Are massive astrophysical BHs described by the Kerr solution of GR?]

\item[$\cdot$ Is there structure close to the BH horizon? ]

\item[$\cdot$ What are the properties of massive BH populations throughout the universe? ]

\item[$\cdot$ How have massive BHs grown over time?]

\item[$\cdot$ What is the nature of the stellar environment and of the accretion disks around SMBHs?]
\end{description}

We present our case as follows. First, in Section \ref{sec:detector} we discuss the technical possibility of constructing a detector to achieve our science goals, and in Section \ref{sec:theories} we briefly summarize the fundamental theories that an improved LISA-type detector could address. Finally, we pass to the science extracted from specific classes of sources of GWs in Sections \ref{sec:emri} and \ref{sec:equal_mass}, discussing inspirals of small objects into SMBH and SMBH mergers respectively.

\section{\detector{}} \label{sec:detector}
\begin{figure}
\begin{center}
	\includegraphics[width=0.84\textwidth]{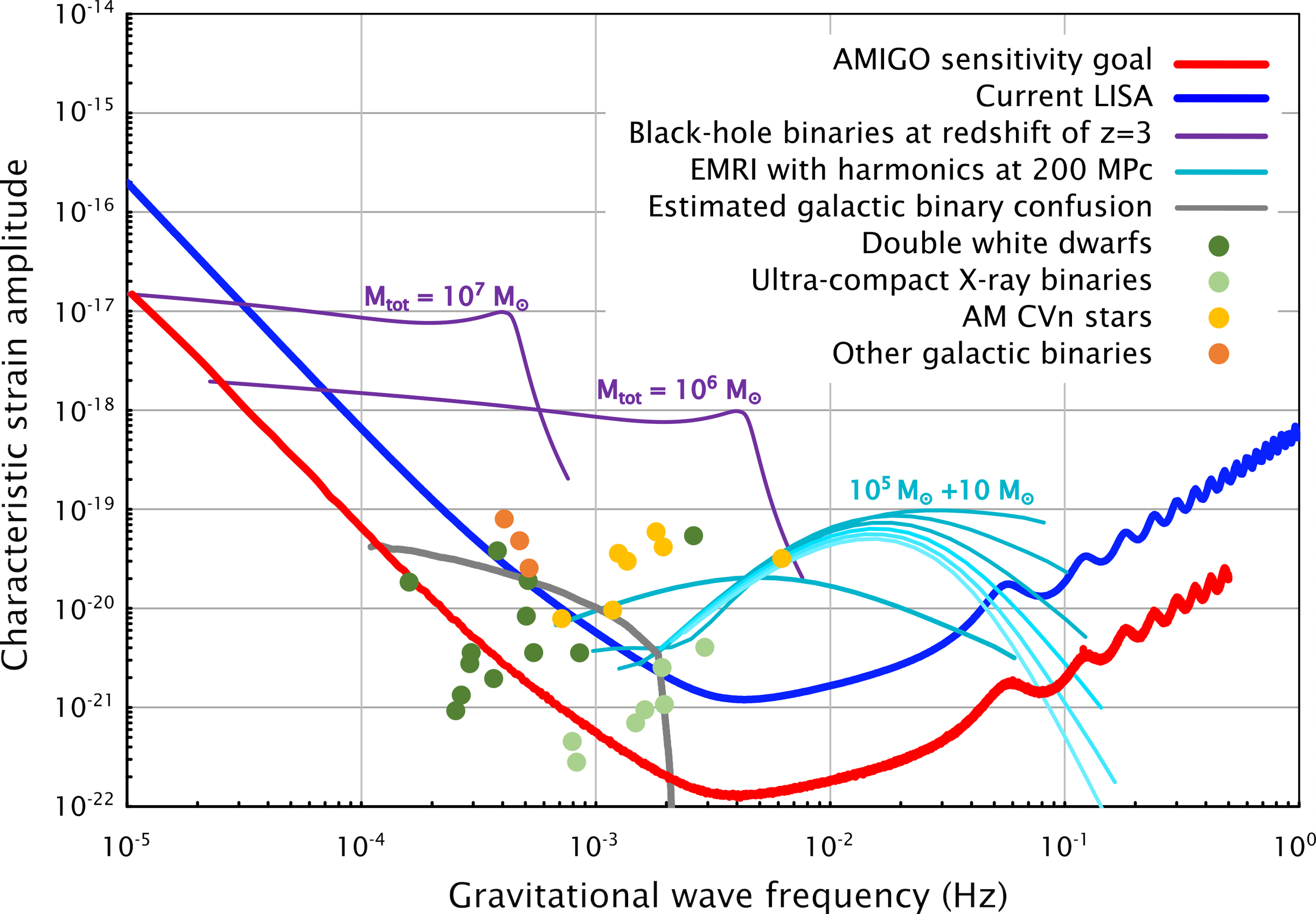}
	\end{center} \vspace{-0.4em}
	\caption{Projected sky-averaged sensitivity curve for \detector{} and possible sources within its range \cite{Cornish:2018dyw}. The sensitivity curve was computed with the use of the LISA Performance Model  (perf-lisa.in2p3.fr) and the sources were adapted from \cite{Barke:2014lsa}.}
	\label{fig:instrument}
\end{figure}

As already mentioned in the Introduction, throughout this White Paper we present a scientific
case for a detector we refer to as \detector{}, an enhanced version of LISA that would improve
its sensitivity in the mHz band by a factor of 10 across all frequencies. An L-class mission fulfilling the AMIGO description could be achieved in the 2035-2050 period under the condition of significant improvements in several key areas compared to the LISA instrument \cite{Audley:2017drz}. 

\paragraph{Shot noise.} The shot noise limited strain sensitivity for an ideal interferometer is:
\begin{equation}
h_{\rm SN} \equiv \frac{\delta L_{\rm SN}}{L} = \frac{\lambda^{3/2}}{D^2}\sqrt{\frac{\hbar c}{\pi P_{\rm in}}}
\end{equation}
which is independent of the interferometer arm length $L$ as long as the laser power $P_{\rm in}$, the telescope diameter $D$ and the laser wavelength $\lambda$ remain the same. Improvements by an order of magnitude can be achieved by increasing the telescope diameter and the laser power and reducing the laser wavelength; most likely a combination of all three. This sensitivity is the target or bench mark for any technology development program dedicated to improvements in the interferometry or more precisely in our displacement sensing system. Limiting noise sources beyond shot noise include stray light, geometric effects that couple spacecraft motion to (apparent) length changes, timing noise in the phasemeter, laser frequency noise, temperature driven optical pathlength changes and many other mostly technical noise sources.

\paragraph{Acceleration noise.} The limiting noise sources at low frequencies in a LISA-like mission are summarized under acceleration noise. The LISA Pathfinder experiments have generated a detailed noise model which can in principle also be used to design an improved mission \cite{Armano:2016bkm,Armano:2018kix}. The low hanging fruit is the reduction in the gas pressure around the test mass which was one of the main limiting noise sources in LISA Pathfinder. Spacecraft motion couples through electro-static actuation forces and through gravitational forces to test mass motion. Larger gaps, better gravitational balancing, $\mu$N thrusters with faster response times, an interferometric readout system that monitors all degrees of freedom with sub-pm sensitivity are all approaches which should enable the necessary improvements.

\paragraph{Galactic Binary Confusion.} In a future mission design, these technological challenges have to be traded against each other but also against the “Galactic Binary Confusion” noise \cite{Nelemans:2001hp} to maximise the scientific pay-off of the AMIGO design. On the other hand, the overlap with the edge of the background of binary sources within the Galaxy would provide a wealth of signals with high signal-to-noise ratios (SNRs) for
calibration, similarly to the case of LISA itself. This future development could make the concept of an
AMIGO detector move closer to that of, e.g., the ALIA mission design \cite{Baker:2019pnp}. 

\paragraph{Detector concept.} As a specific example of a feasible \detector{} concept, we consider an identically placed and spatially configured interferometer as LISA ($2.5$ million km arms), a 0.5 m telescope mirror, and a 30 W laser at the same wavelength as LISA (1064 nm). Furthermore, we assume a 10-fold improvement in acceleration noise as compared to LISA. The resulting noise curve is plotted in Fig. \ref{fig:instrument}. It is interesting to note that this particular configuration would also double the number of resolved galactic binaries (SNR$>$7) and increase the number of mass measurements of the binaries by a factor of order 10.

\section{New theories, old problems} \label{sec:theories}
\detector{} will shed light on some of the key outstanding issues concerning gravitation. These range from unresolved {\it foundational} questions, to how gravity could help shed light on {\it new} fundamental interactions or fields. Before passing to the discussion of concrete sources of gravitational radiation in the \detector{} band, we first present a brief standalone summary of the fundamental theories that are to be confronted with such observations.
\subsection{Testing the cornerstones of GR with BHs and GWs\label{sec:corner}}

Gravitation continues to be the most enigmatic of all interactions. It is poorly understood at the shortest microscopic scales, where quantum effects are important. The need to stipulate the existence of dark matter and dark energy suggests that our understanding of gravity might be questionable at large scales as well. Experimental guidance into the mysteries of gravity is as pertinent as ever, and its impact on fundamental physics and cosmology cannot be overstated. 

Ultimately, one needs to test cornerstones of the theory: are fundamental symmetries, such as Lorentz symmetry or parity invariance, fully respected by gravity? Is the gravitational interaction mediated by the metric only -- in particle physics language, by a massless spin-2 particle called the graviton? Can gravitons be massive? Can there be new fields that partake in the gravitational interaction, either as classical remnants of a quantum description of gravity (e.g. compactification of extra dimensions), or as explanations to dark matter or dark energy?

In practice, alternative theories of gravity can offer quantitative and systematic answers to the questions above. It is straightforward to argue that most deviations from GR (including Lorentz symmetry violations, massive gravitons, extra dimensions, etc.), can effectively be described in the classical regime by a covariant effective field theory with a metric and additional fields~\cite{Sotiriou:2015lxa}.
The nature and the properties of these fields encode the fundamental departure from GR.
Hence, by detecting or constraining alternative theories of gravity and their dynamics, and with the right theoretical underpinning, we are probing the most fundamental principles of gravitation.

\paragraph{Scalars and vectors.} The simplest modification of gravity relies on the introduction of a scalar field. No-hair theorems \cite{Hawking:1972qk,Bekenstein:1995un,Sotiriou:2011dz} imply that scalar fields cannot actually leave an imprint on the structure of quiescent (stationary, asymptotically flat) BHs in theories that do not exhibit derivative (self) interactions. Further investigation \cite{Hui:2012qt, Sotiriou:2013qea} has 
pinned down the unique coupling term between a massless 
scalar and gravity that can give rise to hairy BHs --- a 
coupling with the Gauss-Bonnet invariant. Such hairy BHs 
\cite{Campbell:1991kz,Kanti:1995vq,Yunes:2011we,Sotiriou:2014pfa} 
have received a lot of attention and there is an ongoing effort to explore their dynamics numerically \cite{Benkel:2016rlz,Benkel:2016kcq,Witek:2018dmd}. BHs in general scalar-tensor theories \cite{Horndeski:1974wa} with a massive scalar are less explored. It has recently been demonstrated that curvature couplings can contribute to the effective mass of the scalar, triggering a tachyonic instability and leading to {\em BH scalarization} \cite{Doneva:2017bvd,Silva:2017uqg}: BHs acquire scalar hair only if they lie in a certain mass range. 
Interestingly, no-hair theorems effectively provide guidance for finding hairy BHs, by violating one of their assumptions, such as changing the asymptotics \cite{Jacobson:1999vr}, introducing matter in the vicinity of the BH  \cite{Cardoso:2013fwa}, or considering long-lived instead of stationary configurations, as discussed in detail in Section~\ref{sec_dm}.
Irrespective of its origin, scalar hair alters the structure of the BHs and can generally leave an imprint on any part of GW waveforms (inspiral, merger, ringdown).

One can also introduce an additional vector field into the gravity sector. The most general vector-tensor theories with derivative self-interactions but still second-order equations of motion have been established \cite{Tasinato:2014eka,
      Heisenberg:2014rta,Jimenez:2016isa} and can be viewed as generalised Proca theories with genuinely new vector interactions. 
Again, a generic outcome is that such fundamental fields can be dynamically excited and leave observable imprints on the stationary geometry of BHs~\cite{Chagoya:2016aar,Heisenberg:2018vti}.

\paragraph{Symmetry violations.}  Special classes of theories with additional scalar and/or vector fields are those that violate fundamental symmetries, such as Lorentz invariance or parity invariance. The key property of Lorentz-violating theories is that they exhibit a different causal structure from GR, featuring superluminal or even instantaneous \cite{Blas:2011ni,Bhattacharyya:2015uxt} propagation without causal conundrums. This directly implies that BHs in these theories have to be fundamentally distinct from those of GR, featuring different horizons for excitations travelling at different speeds \cite{Eling:2006ec,Barausse:2011pu,Barausse:2013nwa} and potentially a ``universal'' horizon that blocks even instantaneous propagation \cite{Barausse:2011pu,Blas:2011ni,Bhattacharyya:2015gwa}. This rich causal structure is expected to leave an imprint on waveforms, and modelling BH and binary dynamics in Lorentz-violating theories is an ongoing challenge (e.g.~\cite{Garfinkle:2007bk,Bhattacharyya:2015uxt,Akhoury:2016mrc,Ramos:2018oku,Sarbach:2019yso}). It is worth noting that GWs, and specifically the detection of a NS binary merger (GW170817) with coincident gamma ray emission that constrained the speed of GWs to a part in $10^{15}$ \cite{Monitor:2017mdv}, have already given the strongest known constraint on Lorentz violation in gravity. This is a clear indication of the potential of GWs in constraining fundamental physics. However, it should also be stressed that the speeds of the extra polarizations that Lorentz-violating theories generically exhibit \cite{Sotiriou:2017obf} remain virtually unconstrained even when all known bounds are combined \cite{ Gumrukcuoglu:2017ijh}. This implies that there is still room for sizeable deviations from GR, and that further effort is necessary to rule out Lorentz violations in gravity. 

Similar considerations apply to parity invariance. Dynamical Chern-Simons gravity~\cite{Jackiw:2003pm,Alexander:2009tp} provides an effective description of parity violations in gravity. 
Rotating BH solutions in such theories \cite{Yunes:2009hc,Konno:2009kg,Yagi:2012ya,Stein:2014xba} exhibit (pseudo-)scalar structure outside the horizon and \emph{dipole} ``hair'' that affects the GW waveforms~\cite{Sopuerta:2009iy,Pani:2011xj,Yagi:2012vf,Canizares:2012is,Okounkova:2017yby}.

\paragraph{Multi-metric theories.} Gravity theories with multiple metrics arise naturally from the point of view of extensions of the Standard Model of particle physics, but also from the question of whether gravitons have mass~\cite{deRham:2010kj,Hassan:2011zd}.
The presence and the properties of BH solutions highly depend on the assumptions for the two metrics \cite{Volkov:2012wp,Volkov:2013roa,Brito:2013xaa,Volkov:2014ooa,Babichev:2015xha} and are currently the subject of intense investigation. Current GW bounds on the mass of the graviton rely on propagation effects or dynamical instabilities, and taking into account modification of the binary's dynamics could lead to impressive improvements~\cite{Cardoso:2018zhm}.

To summarize, deviations from GR generically imply the existence of additional fields that mediate gravity. These can leave an imprint on the structure of BHs and also double as additional GW polarizations. The extensive study of scalar-tensor theories has taught us valuable lessons on how to search for new fields and polarizations, and is currently spearheading waveform modelling beyond GR. At the same time, there is an ongoing, systematic exploration of more general deviations from GR and their connections to fundamental physics. This is bringing a leap in our understanding of strong gravity phenomena and BH modelling beyond GR. Maintaining this huge momentum is crucial to be ready to make fundamental physics inferences with LISA and \detector{}.

\subsection{BHs, GWs and dark matter\label{sec_dm}}
Understanding dark matter (DM) is one of the major scientific endeavours of this century~\cite{Bertone:2018xtm,Bertone:2019irm}. Models to explain it range from ultralight bosons with masses $\sim 10^{-22} \mbox{ eV}$ to BHs with tens of solar masses. A considerable effort has been made in the past decades to directly detect new particles that would fit the description of DM, but no experimental evidence of such particles has been found so far. Thus, if DM is made of new fundamental particles, it must interact very feebly with baryonic matter. However even in such a case, Einstein's equivalence principle---one of GR's fundamental pillars---requires that all forms of matter and energy gravitate similarly. Therefore, gravitational probes, such as GWs and BHs, provide an alternative and very compelling laboratory to directly probe the existence of new fundamental particles and to solve the long-standing puzzle of the nature of DM.

\paragraph{DM influence on inspirals.} The GW emission of a binary BH system can be influenced by the presence of DM in different ways. A binary system evolving in a DM-rich environment will be influenced by at least two effects: accretion and dynamical friction (as the bodies move through the DM medium, they accrete material while simultaneously exerting a gravitational pull on the surrounding DM). These effects leave an imprint on the inspiral dynamics and, in consequence, the GW phasing. Although this also occurs for a binary in a baryonic environment (see Section~\ref{sec:environment}), the dynamical friction for collisionless DM and large-scale coherent DM is different from that of normal fluids~\cite{Macedo:2013qea,Macedo:2013jja,Hui:2016ltb}. The magnitudes of accretion and gravitational drag are generically small, but their cumulative effect over a large number of orbits can leave a clear imprint in the gravitational waveform, measurable as a phase shift in the GW signal relative to the inspiral in vacuum~\cite{Macedo:2013qea,Macedo:2013jja,Eda:2013gg,Eda:2014kra,Barausse:2014tra,Barausse:2014pra,Yue:2017iwc,Hannuksela:2018izj,Hannuksela:2019vip}. 

Direct imprints of the DM environment on GW waveforms are expected to be especially important in regions with large overdensities of DM. Such high densities may form around some intermediate mass BHs if they grew adiabatically from small initial seeds~\cite{Gondolo:1999ef,Ullio:2001fb,Bertone:2005xz,Zhao:2005zr,Eda:2013gg,Eda:2014kra,Yue:2017iwc}, or in the presence of ultralight bosons~\cite{Brito:2015oca}. These tests would be especially impressive for extreme-mass-ratio inspirals (EMRIs, see Section~\ref{sec:emri}), in which case the small compact object may perform $\mathcal{O}(10^{5}-10^{7})$ orbits in the band of \detector{}. EMRIs would allow for a very precise mapping of the spacetime and environmental properties around the massive BH. The detection of a DM environment and the measurement of its density can be used to constrain several DM candidates, such as light bosonic or fermionic DM and self-interacting DM~\cite{Eda:2013gg,Eda:2014kra,Hannuksela:2018izj,Hannuksela:2019vip}. With a large number of GW detections, self-interacting DM may also be revealed through measurements of the rate of SMBH mergers as a function of redshift, which will be heavily affected by DM accretion \cite{Markevitch:2003at,Ostriker:1999ee,Hennawi:2001be}.

\paragraph{DM as a GW source.} DM particles around a SMBH may be forced into highly eccentric orbits by the influence of gravitational perturbations from a more massive BH companion~\cite{Naoz:2019pch}. A torus may form near the SMBH, eventually leading to the formation of clumps, driven by the DM self-gravity. If DM self-annihilates, such a clump leads to electromagnetic and GW emission. Depending on the density profile and the mass of the DM particles, DM can accumulate near the ergosphere for a timescale up to $10^8$ yr~\cite{Naoz:2019pch}. During this time, order-of-magnitude estimates of the clump's GW emission indicate that it could be a potential GW source for \detector{}.

 A compelling DM candidate is the QCD axion, or axion-like particles. For such particles, overdensities around BHs can provide new signatures for multi-messenger astronomy. The natural weak interaction of the QCD axion with matter makes its direct detection challenging. However, axions may be efficiently converted to photons in the strong magnetosphere of NSs, potentially observable by current and future radiotelescopes~\cite{Safdi:2018oeu,Hook:2018iia,Huang:2018lxq}. The signal can be significantly enhanced for NSs inspiralling into an intermediate-mass BH surrounded by a dense DM environment. Joint electromagnetic and GW observations of these systems can probe the existence of the QCD axion in the range $[10^{-7},10^{-5}]$ eV and provide very tight constraints on both the DM profile and QCD axion properties~\cite{Edwards:2019tzf}.

\paragraph{New bosons.} In addition, spinning BHs can be unstable against the formation of a bosonic cloud of light fields, whereby a transfer of energy from the BH to the light field occurs via superradiance~\cite{Brito:2015oca}. Bosons with masses in the range $10^{-21}$~--~$10^{-10}$ eV have a Compton wavelength comparable to the radius of astrophysical BHs, and can form efficient bound states. The extracted energy condenses as a bosonic cloud around the BH producing ``gravitational atoms'' with energy up to $\sim 10 \%$ of the mass of the BH~\cite{Arvanitaki:2010sy,Brito:2014wla,East:2017ovw}. The evolution of these systems leads to several observational channels. 

For complex bosonic fields, GW emission is suppressed and the end-state of the instability can be a spinning BH with external bosonic structure~\cite{Herdeiro:2014goa,Herdeiro:2016tmi}. On the other hand for real bosonic fields, the cloud disperses over very long timescales, through the emission of nearly-monochromatic GWs with frequencies that range from $10^{-6}$ Hz up to $10^4$ Hz depending on the boson mass. Such signals can be observable individually or as a very strong stochastic GW background~\cite{Arvanitaki:2016qwi,Baryakhtar:2017ngi,Brito:2017zvb,Brito:2017wnc}. Furthermore, BHs affected by the instability spin down in the process; thus, accurate measurements of the spin of a large population of BHs can strongly constrain, or detect, ultralight bosons in the range $\sim [10^{-21},10^{-10}]$ eV ~\cite{Arvanitaki:2010sy,Pani:2012vp,Pani:2012bp,Brito:2013wya,Arvanitaki:2014wva,Baryakhtar:2017ngi,Brito:2017zvb,Cardoso:2018tly}.

Self-interacting bosonic particles may lead to collapse (``bosenovas''), producing gravitational radiation which may be detectable by \detector~\cite{Yoshino:2012kn,Yoshino:2013ofa,Yoshino:2015nsa} and may even form compact stars in a large range of masses~\cite{Liebling:2012fv}.
The merging of such binaries carries important information on their nature. For example, such objects respond differently to the tidal interaction of their companion~\cite{Mendes:2016vdr,Cardoso:2017cfl,Sennett:2017etc} and would, in general, lead to a different post-merger signal when compared to the merger of two BHs~\cite{Palenzuela:2017kcg,Helfer:2018vtq,Sanchis-Gual:2018oui}. Such effects are imprinted in the gravitational waveform and allows to strongly constrain the nature of the compact object (see Sections~\ref{sec:horizons} and~\ref{sec:equal_mass}).

\paragraph{Charge bounds.} Finally, whereas astrophysical BHs are expected to be electrically neutral, they may carry a hidden electric charge, as suggested by some models of milli-charged DM~\cite{Cardoso:2016olt} and in Einstein-Maxwell-dilaton models~\cite{Julie:2017rpw,Julie:2018lfp,Khalil:2018aaj}. New interactions lead to new emission channels during a binary coalescence, modifying both the orbital dynamics and the post-merger phase of the signal. For loud, nearly equal-mass BH binaries, \detector{} will be able to place exquisite bounds on the corresponding dipolar moment of the system (see Section~\ref{sec:equal_mass}).

\subsection{BHs and the birth of our universe}
\detector{} will also inform us on how the early universe looked. The same inflationary mechanism that generates large scale fluctuations in the cosmic microwave background and the large scale structure of the universe (galaxies and clusters) also generates fluctuations on small scales. Some models of inflation predict that large curvature fluctuations were generated thirty e-folds before the end of inflation, and then collapsed upon horizon reentry to form BHs during the radiation era~\cite{GarciaBellido:1996qt}. These primordial BHs (PBHs) are formed from photons which cannot escape gravitational collapse of large curvature perturbations generated from quantum fluctuations during inflation. They have masses in the range between planets ($10^{-6}\,M_{\odot}$) and SMBHs ($10^{6}\,M_{\odot}$), and may form due to sudden changes in the relativistic fluid’s radiation pressure, as particles decouple or condense throughout the thermal history of the Universe~\cite{Hawking:1971ei,Carr:2019kxo}. 

In the matter era, these massive BHs act as seeds for structure formation~\cite{Clesse:2015wea}. They could form disks very early on and grow via gas accretion to gigantic sizes ($10^9-10^{11}\,M_{\odot}$) in a Hubble time. These seeds could be responsible for the gradual (rather than explosive) and uniform (rather than patchy) reionisation of the Universe~\cite{Garcia-Bellido:2017fdg}. Their growth occurs mainly through gas accretion, but there is a possibility that they merge to form massive clusters of PBHs with a wide mass distribution, segregated in mass with the most massive at the center due to dynamical friction, and the least massive orbiting or even evaporating from these clusters and thus constituting a diffuse uniform component of dark matter. In that case, one expects a large rate of hyperbolic encounters as well as mergers with extreme mass ratios, at high redshifts~\cite{Ali-Haimoud:2017rtz}. 
The spatial distribution and mass range of PBH constitute a specific signature for microlensing surveys and GW events~\cite{Clesse:2016vqa}.

\subsection{BHs and horizons\label{sec:horizons}}
Finally, \detector{} will provide an exclusive glimpse of the truly unique prediction of GR: BHs. As was mentioned, BHs in GR are very simple objects, but characterized by a special boundary: horizons, the defining feature of BHs in classical GR. Horizons constitute a causal boundary; information can propagate from the exterior universe to the BH interior, but at the classical level nothing can cross in the other direction. Once quantum effects are included at the semiclassical level, BHs appear to evaporate thermally. The ``information paradox'' resulting from this process~\cite{Hawking:1976ra} suggests that the classical picture of BHs is at least incomplete~\cite{Unruh:2017uaw,Compere:2019ssx}. A resolution likely requires changes at the horizon scale, which can even be drastic to the point of predicting new, very compact but horizonless objects in the universe~\cite{Mathur:2005zp,Bena:2007kg,Bena:2016ypk,Mazur:2004fk,Barcelo:2007yk,Almheiri:2012rt,Giddings:2013kcj,Giddings:2017mym,Carballo-Rubio:2017tlh,Berthiere:2017tms,Kaplan:2018dqx}.
Thus, BHs and NSs might be only two ``species'' of a larger family of astrophysical compact objects~\cite{Cardoso:2019rvt}. More exotic objects (that we will call exotic compact objects, or ECOs) are also predicted in modified theories of gravity, or when GR is coupled to exotic matter or beyond-Standard-Model fields (cf. Sections~\ref{sec:corner} and \ref{sec_dm}).

From a fundamental gravity standpoint, BHs and their horizons offer a unique possibility to examine one of the most extreme and important predictions of GR. Dynamical horizons are not only predictions of gravitational collapse in GR: their existence is {\it necessary} for the consistency of the classical theory. This is the root of Penrose's Cosmic Censorship Conjecture~\cite{Penrose:1969pc,Wald:1997wa}, which remains one of the most urgent open problems in fundamental physics. The statement that there is a horizon in any spacetime harboring a singularity in its interior is a remarkable claim -- and it requires remarkable evidence.

In standard GR, BH horizons have properties which influence binary systems and the GWs that they generate. Because a horizon is a causal boundary, fields must satisfy purely-ingoing boundary conditions there. The horizon is in addition a surface of infinite redshift. If the supermassive dark objects observed at the centers of galaxies and expected to form during galaxy mergers differ from the standard BH picture, these differences could leave an observable imprint on the GW signal from both isolated and binary dark compact objects.

How well can observations probe these properties? The answer to this question depends on the particular effect under study and the specific GW source. We list several promising possibilities here, remarking on the prospects for observing them.

\paragraph{Multipole moments.} 
Generally ECOs have a multipole moment structure which differs from that of BHs. The number of independent moments is larger and the Kerr symmetry properties can be broken~\cite{Ryan:1995wh,Vigeland:2009pr,Raposo:2018xkf}. These moments affect the dynamics of the inspiral, leaving an observable imprint on the emitted GWs~\cite{Blanchet:2006zz}. Deviations from the standard BH multipole moments can be found by measuring the inspiral phase of a comparable-mass binary coalescence~\cite{Krishnendu:2017shb,Kastha:2018bcr}, or by mapping the spacetime geometry of supermassive objects as probed by an EMRI~\cite{Ryan:1995wh,Barack:2006pq,Babak:2017tow}: in Sections \ref{sec:emri} and \ref{sec:equal_mass} below we show just how loud these events can be for \detector{}. \detector{} will set impressive bounds on the quadrupole moment of SMBHs with EMRIs, while also making it possible to cleanly measure moments beyond mass and spin for comparable-mass binaries.

\paragraph{Tidal heating.} Each member of a binary feels the other's tidal field. These tides backreact on the orbit, transferring energy and angular momentum from their spin into the orbit.  This effect is called {\it tidal heating}: for fluid bodies, each member is heated by the tide's shear on its fluid. Tidal fields on BHs satisfy a unique boundary condition which picks out how a BH's spin is transferred to the orbit. This effect can be responsible for thousands of radians of accumulated orbital phase for EMRIs in the band of a space-based GW antenna~\cite{Hughes:2001jr,Datta:2019euh}. This is due to the dissipative nature of horizons, which make BHs perfect absorbers. If at least one binary member is an ECO, the dissipation is likely to be much smaller, even negligible, potentially changing the inspiral phase by a large amount (especially if the binary's members spin rapidly).
Extending the results of Ref.~\cite{Maselli:2017cmm} to \detector{}, we estimate that SMBH coalescences up to cosmological distances ($z\sim 10$) can be 
confidently distinguished from ECO binaries by the presence of tidal heating.

\paragraph{Tidal deformability.} The nature of inspiraling objects is also encoded in how their shapes respond to the tidal field of a companion.  This {\it tidal deformability} is quantified by their tidal Love numbers~(TLNs)~\cite{PoissonWill}. A remarkable result of GR is that the TLNs of all (slowly spinning) BHs are exactly zero~\cite{Damour_tidal,Binnington:2009bb,Damour:2009vw,Gurlebeck:2015xpa,Pani:2015hfa,Pani:2015nua, Landry:2015zfa}, while those of compact horizonless objects are small but finite~\cite{Pani:2015tga,Uchikata:2016qku,Porto:2016zng,Cardoso:2017cfl,Giddings:2019ujs}. This precise cancellation suggests a powerful test: a non-vanishing TLN with measurement errors small enough to exclude zero provides a smoking gun for ECOs.  In some models, TLNs can depend logarithmically on an object compactness, near the BH limit~\cite{Maselli:2017cmm,Cardoso:2019rvt}. This poses a challenge: the errors on the compactness are exponentially sensitive to the TLNs~\cite{Addazi:2018uhd}.  The outlook for constraining this physics is marginal with LISA, but \detector{} would provide the opportunity to measure TLNs of  extreme models, possibly even placing bounds on a putative effective surface at Planckian distance from the would-be horizon~\cite{Addazi:2018uhd}. Prospects to measure the TLNs of supermassive objects are much better with EMRIs~\cite{LoveExtreme}, and would significantly improve with \detector{}, constraining this physics several orders of magnitude better than current limits from LIGO/Virgo for NSs.

\paragraph{Mergers of dark objects.} Simulations of ECO coalescences and mergers have so far only been done for boson stars (see~\cite{Liebling:2012fv} for a review). Several distinctive features have been found, such as the excitation of post-merger scalar modes, the ejection of scalar ``blobs,'' and the formation of a non-spinning boson star as the remnant of a moderately compact binary~\cite{Palenzuela:2017kcg}.  These studies show that GW waveform differs significantly from BH models, at least for the cases examined to date~\cite{Palenzuela:2017kcg}, demonstrating the potential of waveform consistency checks to tests the nature of merging systems~\cite{TheLIGOScientific:2016src}. Numerical simulations in other ECO models are needed to extend such tests. 

\paragraph{Ringdown and post-ringdown spectroscopy.} If an ECO remains after merger, the post-merger phase is particularly rich.  Matter and fluid modes may be excited; some modes may be resonantly excited.  The remnant's quasinormal mode spectrum differs from that of a BH, providing a further method for testing BH physics~\cite{Berti:2005ys}. The boundary conditions of ECOs are dramatically different from those of a BH, which can lead to ``echoes'' of the prompt ringdown~\cite{Cardoso:2016rao,Cardoso:2016oxy,Oshita:2018fqu,Wang:2019rcf,Kokkotas:1995av,Ferrari:2000sr,Pani:2018flj,Buoninfante:2019teo,Delhom:2019btt}; see~\cite{Cardoso:2017cqb,Cardoso:2019rvt} for recent reviews.
Echoes in the GW data would allow us to probe the near-horizon structure of compact objects.  The absence of echoes could place increasingly stronger constraints on alternatives to the BH paradigm, for example on the effective reflectivity of the object~\cite{Maggio:2019zyv}. An improvement in sensitivity by a factor $10$ relative to LISA will allow us to put constraints on the ECO reflectivity at $5\sigma$ confidence level, which are impossible with near-future detectors.

\paragraph{Bottom line.} 
Future GW instruments in the mHz band can be used as a probe of the BH nature of heavy and dark objects, but also to test the putative signatures of alternative proposals and to exclude (or discover) supermassive dark objects other than BHs.  An improvement in sensitivity by a factor of at least $10$ relative to LISA will be crucial in this context.

Most of the above questions require a sensitive GW-detector such as \detector{}. They also require clean astrophysics and fundamental-physics probes, and BH systems are ideally placed for such a role. The following is a concise overview of the potential of two BH-systems that we know exist in our universe and that will be probed by \detector{}.
\section{EMRIs: ideal probes of black-hole physics\label{sec:emri}}

Within the class of compact-object binary systems, the ubiquity of SMBHs residing in galactic nuclei has made binaries with a mass ratio $10^{-5}$ -- $10^{-7}$ a prime target for low-frequency space-based GW observatories. 
These systems we have already referred to as EMRIs, binary systems in which one component is substantially more massive than the other, hold the exciting promise of providing unrivalled tests of Einstein's GR. They also hold the promise of improving our understanding of the properties of galactic nuclei, and of enabling precision measurements of the properties of SMBHs out to high redshift.

\begin{figure}
\begin{center}
	\includegraphics[width=8cm]{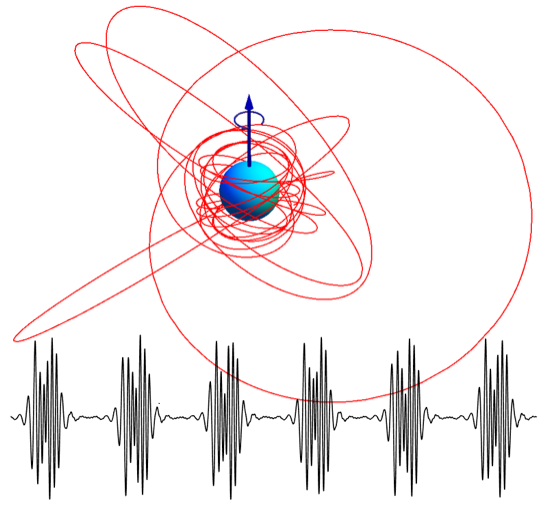}
	\caption{The complicated trajectory of a compact object inspiralling into a SMBH (red curve) leads to richly structured GWs (bottom, black curve) which carry with them a wealth of information about the binary and the environment in which it lives.}\label{fig:EMRI}
\end{center}
\end{figure}

The EMRI regime is qualitatively very different from the comparable mass regime where ground-based detectors have been so successful \cite{LIGOScientific:2018mvr}.
The small mass ratio of EMRIs leads to a slower evolution of the binary.
As such, the light secondary completes many tens of thousands of orbits at relativistic speeds in the strong gravitational field of the heavy primary. 
Furthermore, the binary tends not to have completely circularized, resulting in orbits with a very rich structure (see Fig. \ref{fig:EMRI}). 
The resulting GWs carry with them exquisite information about the spacetime of the binary and the environment in which the EMRI lives. 
Regarding the latter, there remains a great deal of uncertainty about the nature of the stellar environment around galactic nuclei.
As we discuss below, the \detector{} mission will start to fill in this gap in our knowledge by enabling a wide range of studies of the stellar environment around SMBHs. 
In addition to EMRIs, \detector{} will also search for intermediate mass-ratio inspirals (IMRIs) with a mass-ratio in the range $10^{-2}$ -- $10^{-3}\!$. 

In order to extract the maximum science gain from EMRI and IMRI observations it will be critical for a next-generation space-based GW mission to have: (i) high signal-to-noise ratios; (ii) a sufficient number of sources to draw statistically significant conclusions; and (iii) a large-redshift observational horizon distance. These three features will provide a unique opportunity to answer key scientific questions, such as: how have SMBHs grown over cosmic time? what is the nature of the dense stellar environment around SMBHs? what is the fundamental nature of gravity in the extreme region near a BH event horizon? A high-sensitivity millihertz GW detector would be ideally positioned to realise these goals and answer these key scientific questions. In the following subsections we address each of these questions in turn.

\subsection{How have SMBHs grown over cosmic time?}
An \detector{} mission would be sensitive to EMRI and IMRI events to great
cosmological distances, probing events that involve the earliest BHs to form in our universe.  In the redshifted total mass range
$10^4\,M_\odot \lesssim (1 + z)M \lesssim 10^7\,M_\odot$, \detector{} would
extend the high precision studies of BH properties that we
expect from LISA from $z \sim 3$ to $z \sim 10$ or even more.  Such
redshifts are comparable to, or larger than, those of the earliest known quasars \cite{Banados:2017unc}. In order to power the high-redshift
quasars we see today, massive ($M \gtrsim 10^8\,M_\odot$) BHs
must exist, but there is great uncertainty regarding how they formed
and grew to these masses. Simulations of early galaxy and structure
growth show that galaxies and galaxy halos grow through the repeated
merger of smaller-scale structures, often at mass ratios of 20:1 or more
(see \cite{Natarajan:2019ipd} and references therein for discussion). If
these structures host BHs whose masses scale with the
structure's mass (an assumption that is approximately consistent with BHs observed in the local universe \cite{McLure:2001uf}), then
the assembly of the first SMBHs would be accompanied by
many EMRI and IMRI events.

Observations of high-redshift EMRIs and IMRIs would provide
sub-percent-level accurate determinations of SMBH masses, and
determine their spins with roughly the same level of
precision.  By measuring a large population of such events, \detector{}
could track the co-evolution of BH mass and spin from very
early times, revealing whether there are trends in massive BH properties as the universe evolves from its earliest moments. An additional intriguing prospect concerns the measurement of peculiar velocities using \detector{}: when a binary orbits a SMBH, the emitted GWs have a small Doppler shift which can be measurable and provide precious information on the system and its host galaxy~\cite{Tamanini:2019usx,Wong:2019hsq,Inayoshi:2017hgw,Bonvin:2016qxr}. The mass of the smaller member of the binary would also be precisely determined, teaching us about the properties of the earliest BHs.  These BHs would likely be remnants of the first stars to form in the universe~\cite{2016MNRAS.457.3356V}, making possible direct observational inferences about these unusual and important objects.

The great reach of \detector{} to EMRI and IMRI events means that there will be a challenge of source confusion.  The rate of EMRI events in LISA is fairly uncertain, ranging from several events per year to hundreds or even thousands. Even if LISA's rate is on the low side, the rate for \detector{} is likely to be at least several hundred events
per year.  These events will all be ``on'' simultaneously.  Careful analysis methods will need to be developed in order to disentangle these events from each other.

\subsection{What is the nature of the dense stellar environment around SMBHs?}
\label{sec:environment}
EMRI waveforms will carry a wealth of data about the binary and the stellar environment in which the EMRI formed. This can include information about third bodies that are sufficiently close or heavy,  properties of the secondary, or the presence of an accretion disk. 

\paragraph{External halos and large-scale fields.} Conservative estimates for the {\em gravitational} perturbations caused by dark matter, molecular-gas or stellar halos, accretion disks and various other fields suggest that these are negligible for the purpose of detection and parameter estimation in EMRIs \cite{Barausse:2014tra,Barausse:2014pra}. However, conservative estimates show that the gravitational effects of outstanding perturbing bodies will be observable by \detector{}, as we discuss below. In non-conservative scenarios where, for instance, compact rings or disks of masses and sizes comparable to the central BH perturb the motion, the effects on the inspirals can be large \cite{Barausse:2006vt,Sukova:2013jxa}. A common feature of the inspirals in such strongly perturbed systems would be prolonged resonant episodes and even chaos  \cite{Gair:2007kr,LukesGerakopoulos:2010rc}. 

\paragraph{Perturbing bodies.} Two important scenarios of perturbing bodies external to the binary are: 
(i) a massive BH within several tenths of a parsec of the EMRI \cite{Yang:2017aht,Yunes:2010sm} and 
(ii) stellar-mass objects with $M \sim0.1$-$100 M_\odot$ at several AU from the EMRI \cite{Bonga:2019ycj}. 
In the first scenario, the presence of the massive BH changes the location of the innermost stable circular orbit significantly and thereby the time of merger.
Additionally, the nearby massive BH accelerates the EMRI and induces a small phase drift in the GWs. 
Detecting this phase drift would give a measurement of the ratio of the perturber’s mass to its distance cubed.  
The degeneracy between the mass and distance of the perturber can be broken if the perturber is very close (within a few tenths of a parsec) so that additional derivatives of the motion can be measured.
Such measurements of the mass and distance of potentially nearby massive BHs can help constrain models of galactic merger dynamics.

In the second scenario, the dominant effect of nearby stellar-mass objects on the waveform is through tidal resonance, which can be thought of as the general relativistic version of the Kozai-Lidov resonance. 
These types of resonances are transient but frequent, and they can alter the waveform significantly. 
Observing these resonances constrains the ratio of mass over distance cubed of the nearby perturbers. 
By observing these ``outliers'' close to the central BH, one can constrain stellar mass population models. The likelihood of observing tidal resonances increases significantly with the increased detection rate of \detector{}.

\paragraph{Inspiralling binaries.} Another interesting possibility are binary-EMRIs (b-EMRIs) where a massive BH captures a stellar-mass BH binary. 
This scenario could occur in as many as 10\% of EMRIs \cite{Addison:2015bpa,Chen:2018axp}. 
If the stellar-mass binary merges close to the massive BH, it leaves a unique signature in the waveform from which one can determine 
(i) the recoil velocity of the merged remnant to a precision of 10 km/s (a much higher precision than the 120-200 km/s presented in \cite{Gerosa:2016vip,CalderonBustillo:2019wwe}) and 
(ii) constrain the rest mass lost in the form of GW radiation to an accuracy of 1.5\%. 
Additionally, the merger of the stellar-mass binary produces high-frequency waves detectable by ground-based GW observatories. 
Using this multi-band information, one can improve the current constraints on the graviton mass by an order of magnitude \cite{Chen:2018axp}. 
Given that only 30\% of b-EMRIs are expected to merge close to the central BH, the likelihood of observing such systems and the exciting physics one can extract from them will be significantly higher with the increased range of \detector{}.

\paragraph{Accretion disks.} An \detector{} mission provides also an exciting opportunity to learn about accretion physics in active galactic nuclei (AGNs).  
Most galactic nuclei in the universe are quiescent, especially at $z \lesssim 2$.  
Massive BHs in quiescent nuclei are expected to be surrounded by thick radiatively inefficient gas flows, whose influence on an EMRI is typically negligible \cite{narayan,bcp2014}.
Most EMRIs observed with LISA are thus unlikely to be influenced by accretion.
With \detector{} increasing the range of EMRIs (and IMRIs) to $z \gtrsim 10$, the likelihood of observing an EMRI from an AGN increases substantially. 
In AGN disks, the effects of accretion, dynamical friction and planetary migration on the EMRI waveform are important -- typically more important than GW emission at separations $>20$ -- $40$ gravitational radii, and are stronger than second-order gravitational self-force effects at any separation \cite{Kocsis:2011dr,Barausse:2014tra,Barausse:2014pra}.  
The large uncertainties in accretion disk physics will make precise modeling of EMRIs interacting strongly with disks very challenging. 
As such the higher SNR of \detector{} will be beneficial as it will allow the GWs from the brightest of these systems to be extracted from the data without using template methods.

Since oscillation frequencies of accretion disks are often very close to the orbital frequencies \cite{Blaes:2006fv,Mishra:2018zht}, they are likely to be excited by the passages of a compact object in an IMRI or an EMRI with a corresponding electromagnetic counterpart. The increased sensitivity of the  \detector{} design would allow to detect EMRIs and IMRIs early enough to launch parallel X-ray observation campaigns \cite{McWilliams:2009bg,McWilliams:2011zs}, which would search for associated phenomena such as the elusive quasi-periodic oscillations in the X-ray flux of the host AGN \cite{Vaughan:2005nw,Middleton:2011mf}. This multi-messenger observation would be an invaluable source of data not only for accretion-disk modeling and BH physics, but also a standard siren for cosmology~\cite{Abbott:2017xzu}.

\paragraph{Properties of the secondary.} The secondary in an EMRI is expected to be captured from the central dense stellar environment about which little is known. 
Thus by studying the nature of the secondary itself, information can be gleaned about this environment. 
In particular, we can study the multipole moments of the secondary, including its mass, spin and mass quadrupole moment. 
The latter two introduce very small deviations from the geodesic motion, scaling with powers of the mass ratio.  
However, due to the long duration of the inspiral they can accumulate to give a measurable phase shift in the waveforms.  
In the case where the inspiralling object is compact, like a BH or NS, we can deduce a component of the spin \cite{Witzany:2019dii,Witzany:2019nml} from the respective waveforms.  
Some quadrupole terms could be measured for less compact objects \cite{Vines:2016qwa}. Another effect that the spin and higher multipole moments introduce into the inspiral are prolonged resonances \cite{zelenka2019dynamics}, whose effect, however, requires more study \cite{Ruangsri:2015cvg}. 
The improved sensitivity of \detector{} would allow for improved estimations of the multipoles of the secondary object and of the impact of these resonances on the inspiral.

Some of the most exciting opportunities to study the nature of the secondary would come from electromagnetic counterparts. \detector{} could obtain large numbers of GW signals correlated with the multi-band  tidal disruption events of white dwarfs in IMRIs \cite{Sesana:2008zc}.

\paragraph{Environment of Sgr A*.} Finally we mention the potential to study the environment around our own Milky Way's SMBH, Sgr A*. Theoretical models suggest that there could be a dark cluster of compact objects and low-mass stars within 0.04 parsecs of Sgr A* \cite{Alexander2017,Amaro-Seoane:2019umn}. Fortuitously, the mass of Sgr A* is such that the frequency of GWs from objects within a few hundred gravitational radii are in the most sensitive part of the \detector{} noise curve \cite{Freitag:2002nm,Gourgoulhon:2019iyu}. Searching for objects in this dark cluster will be complementary to ongoing electromagnetic searches for low-mass stars in this region \cite{GC_Astro2020}.
\subsection{What is the fundamental nature of SMBHs?}
As EMRIs spend many thousands of cycles orbiting in a strong gravitational field, the phasing of the GWs will be highly sensitive to the structure of the spacetime near the SMBH. 
This will enable us to answer key question such as ``is the spacetime around a SMBH described by the Kerr metric of GR?'' \cite{Vigeland:2009pr}. 
The improved sensitivity of the \detector{} mission translates to a commensurate improvement over the already excellent, percent-level bounds that will be placed by the LISA mission \cite{Barack:2006pq,Babak:2017tow}. 
It will also open up possible new tests.

 By a rough estimate, the ratio of SNR in the ringing to SNR in the inspiral is $\sim\sqrt{Qm/M}$, where M/m is the large mass ratio, and Q is the “quality factor” of the dominant ringing mode (roughly the number of ringing oscillations).  For most BHs, we expect $Q$ in the range $10 -- 20$.

\paragraph{Ringdown.} The final cycles of any coalescence involving BHs is a {\it ringdown}, the characteristic ``shaking'' of the remnant BH as it settles down to the Kerr solution of GR. These ringdown waves, known as quasinormal modes (QNMs), encode the mass and spin of the remnant BH. If multiple ringdown modes can be measured, the waves can be used to test the hypothesis that the Kerr solution describes this BH, and allows one to learn about the geometric properties of the binary that generated the GW signal.
For essentially all EMRI events, the ringdown will not be measurable even with \detector{}, though it may be accessible for IMRI events. By a rough estimate, the ratio of SNR in the ringing to SNR in the inspiral
is $\sim\sqrt{Qm/M}$, where $M/m$ is the large mass ratio, and $Q$ is the ``quality factor'' of the dominant ringing mode (roughly the number of ringing oscillations). For most BHs, we expect $Q$ in the range $10 - 20$. For typical EMRIs this SNR ratio works out at the $\sim 1\%$ level, so the ringdown's SNR is negligible except for extremely loud EMRIs.  However, for IMRIs with $N_{\rm cyc} \lesssim 10^3$, the ringdown SNR could be within an order of magnitude of that for the inspiral.  Measuring this final
ringdown will provide a powerful check on our understanding of these systems, probing the nature of the final spacetime and the binary's properties in a manner that is completely independent of the techniques used to measure the preceding inspiral.

\paragraph{Near-extremal Kerr BHs.}
The higher EMRI event rate of \detector{} will greatly increase the likelihood of observing a so-called near-extremal BH. The cosmic censorship conjecture implies a fundamental bound on the angular momentum $J$ of BHs in terms of the mass, $J < M^2$. BHs that nearly saturate this bound are called near-extremal. 

Accretion is the main known mechanism to spin up BHs. In geometrically thin disk models, spin can be increased up to the Thorne bound $J/M^2 \approx 0.998$ \cite{1974ApJ...191..507T}.
This bound is reached in SMBH evolution models with mass $10^9 M_\odot$ and higher, assuming anisotropic accretion \cite{Berti:2008af}. In the presence of magnetic fields, the Thorne bound can be by-passed~\cite{Sadowski:2011ka,Gammie:2003qi}, and, generally, there is no fundamental limitation on how close to near-extremality a BH can be.
Inspiral and plunge of a compact object into a supermassive near-extremal BH is therefore a hypothetical but candidate rare event that could be potentially observed by a thorough survey of EMRIs and IMRIs with \detector{}.
The discovery of such a near-extremal BH would be fascinating as these objects are tetering on the edge of becoming naked singularities and a whole plethora of new physics occurs in their near-horizon regime \cite{Marolf:2010nd,Porfyriadis:2014fja, Yang:2013uba,Compere:2017zkn}, a region which EMRIs are ideally suited to probe \cite{Gralla:2016qfw, Compere:2017hsi}. 

The specific mathematical properties of near-extremal black holes lead to the development of ``smoking-gun'' signatures in EMRIs. Specifically, EMRIs develop exponentially decaying tails at fixed oscillation frequencies and with suppressed amplitudes as $(1-J^2/M^4)^{1/6}$ \cite{Porfyriadis:2014fja,Gralla:2016qfw}. Additionally, the plunging phase leads to prolonged ringdowns that exhibit effectively power-law decays due to the overstacking of quasinormal mode overtones \cite{Yang:2013uba,Compere:2017hsi}. The discovery of a coalescence involving a near-extremal BH would therefore provide new qualitative tests of GR in the strong-field regime.

\section{Collisions of supermassive black holes: the loudest events in the universe\label{sec:equal_mass}}
\begin{figure}[htp]
\begin{center}
 \includegraphics[width=0.9\textwidth]{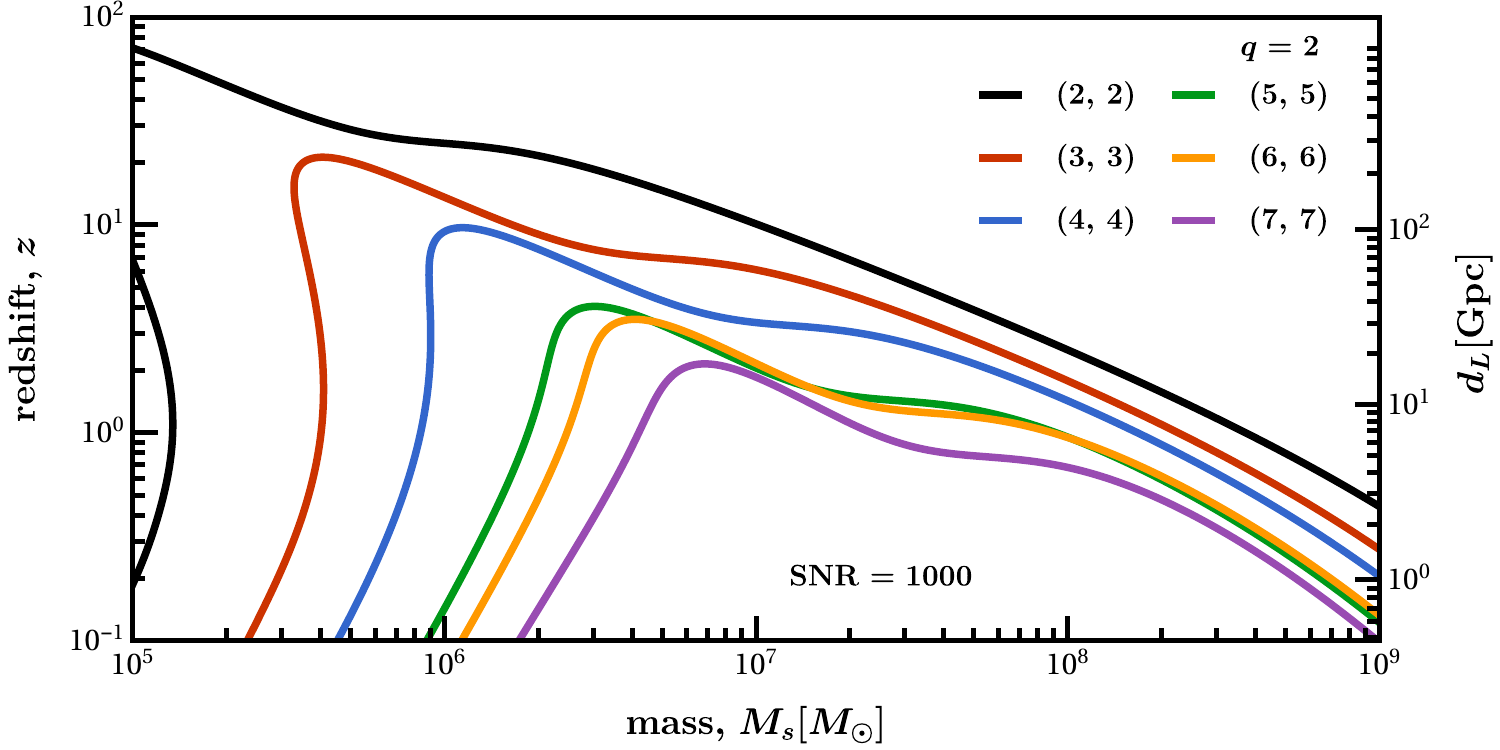}
\end{center}
  \caption{Redshift $z$ and luminosity distance $d_L$ as a function of the remnant BH mass for which \detector{} can see the fundamental mode of different ringdown angular harmonics with an optimal ${\rm SNR}=1000$ for nonspinning BH binary merger with mass ratio $q=2$. SNRs were calculated assuming a ten-fold reduction in the LISA noise and following Ref.~\cite{Baibhav:2017jhs} to calculate the ringdown energy.} 
\label{fig:RDHorizon}
\end{figure}
\begin{figure}[ht]
\begin{tabular}{cc}
  \includegraphics[width=0.45\textwidth]{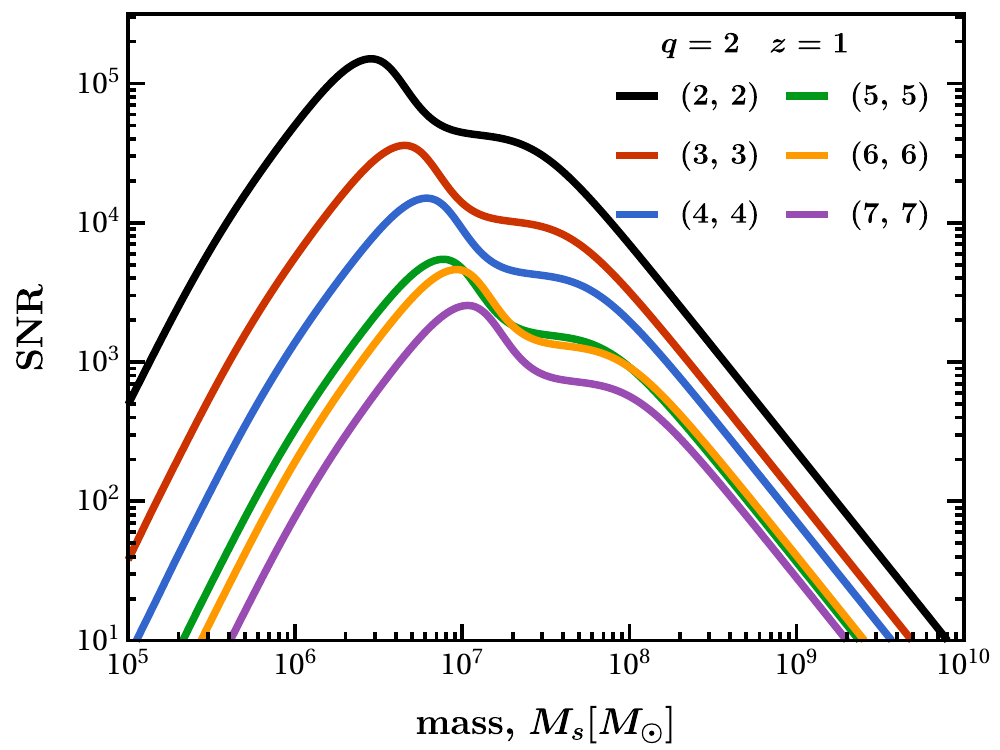}& 
  \includegraphics[width=0.45\textwidth]{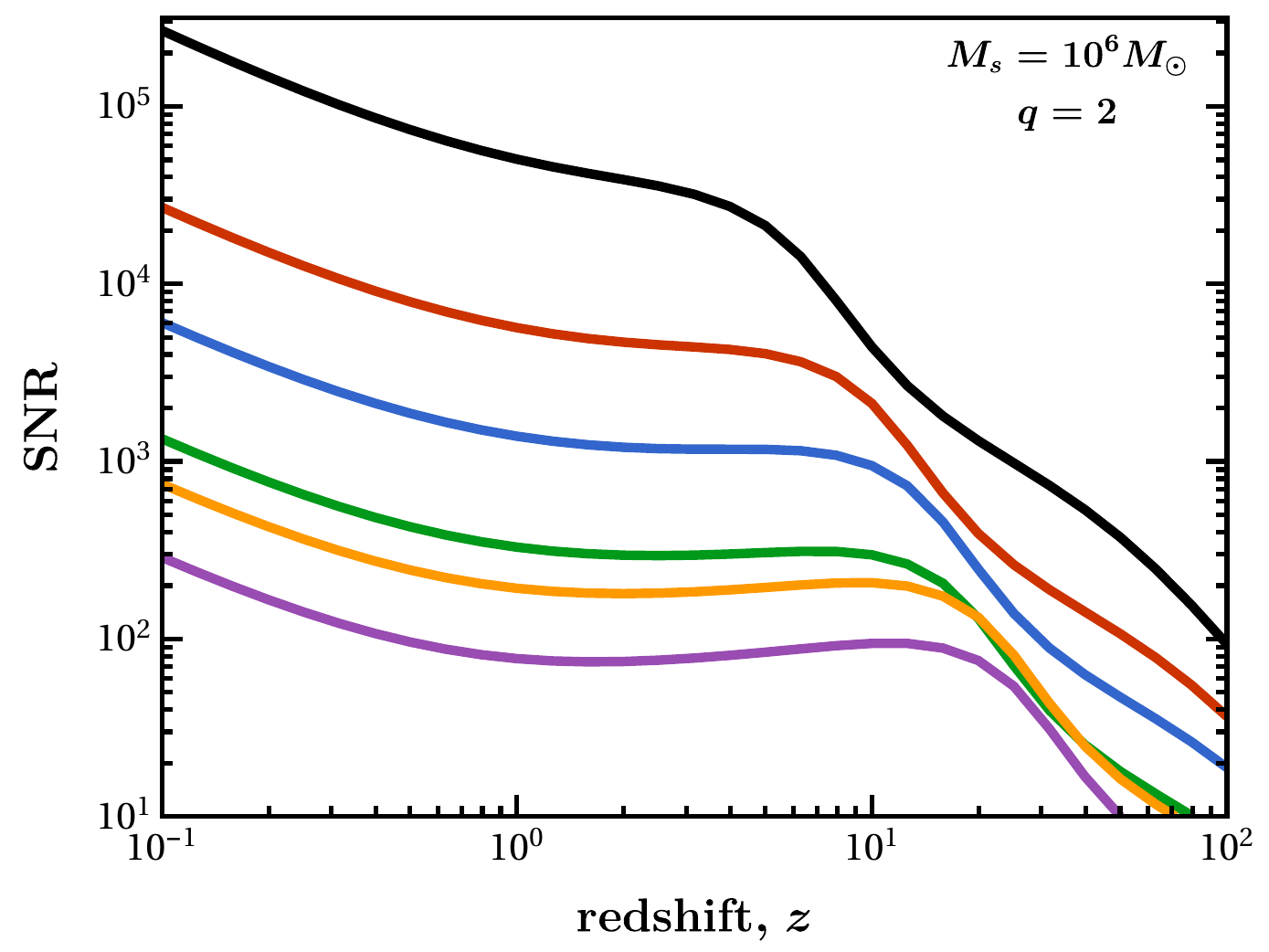} 
\end{tabular}
  \caption{SNR as a function of the remnant BH mass (left) and redshift (right) for the fundamental mode of different ringdown angular harmonics for an optimally oriented, nonspinning BH binary merger with mass ratio $q=2$.} 
\label{fig:SNR}
\end{figure}

As already stated several times, the centers of most galaxies harbour SMBHs. Following a galactic collision, the merger of these BHs generates copious amounts of GWs. 
Such events will be observed by LISA with typical SNRs above 1000 if they occur after the peak of star formation (i.e., at $z\lesssim 2$), and they can be detected (SNR$\gtrsim10$) even beyond $z\sim 20$. The GW signal can be correlated with electromagnetic counterparts from ensuing tidal disruption events \cite{Stone:2010sr}. A mission with one order of magnitude improvement in sensitivity across the LISA band will detect events with BH masses spanning $10^2$--$10^{10} M_\odot$ and possible ${\rm SNRs} \gtrsim 10^4$ (see Fig.~\ref{fig:RDHorizon}--\ref{fig:SNR}), enabling exquisite probes of the properties of BHs and tests of fundamental physics that are simply out of reach for LISA. For systems that emit at the edges of LISA's frequency band, such a mission may increase detection rates by a factor of $1000$. This includes GWs from the most massive SMBHs, currently probed only by pulsar timing arrays, and stellar-origin or intermediate-mass BH binaries, which may also be detected by ground-based GW detectors as multiband systems~\cite{Sesana:2016ljz}. Thus, \detector{} will bridge the entire GW spectrum, enabling tests of GR across 16 orders of magnitude of spacetime curvature.

\subsection{The inspiral of SMBHs: unprecedented tests of gravity}
The analysis of SMBH binaries divides naturally into the long inspiral phase and the violent merger and ringdown phase.
Space-based detectors will observe long inspirals of SMBHs, thereby surpassing any previous tests of the nature of BHs. For mergers of $10^{5}-10^{7} M_\odot$ mass BHs at $z=1$, \detector{} could measure chirp mass and symmetric mass ratio with fractional errors as low as $10^{-7}$ and $10^{-5}$~\cite{Berti:2004bd}. 

\vspace{2pt}
\paragraph{Parametrized tests.}
GR makes highly specific predictions for the GWs emitted by two coalescing BHs. As a result, precise measurements of the phasing of observed GWs could unveil a wide range of departures from the expected behavior of astrophysical BHs in GR. Generic constraints can be placed by explicitly parametrizing deviations from GR in the waveform \cite{Arun:2006yw,Yunes:2009ke}, as has already been done with ground-based detectors \cite{TheLIGOScientific:2016src,LIGOScientific:2019fpa}. 
Each parameter that deviates by a measurable amount from GR could provide evidence for new physics. 

A salient example of such a possible deviation is the emission of dipolar radiation during the inspiral~\cite{Barausse:2016eii,Cardoso:2016olt}. This would be indicative of energy loss through a non-gravitational interaction, due to the BHs possessing electric or, more exotically, some kind of scalar charge \cite{Berti:2013gfa,Barausse:2016eii,Cardoso:2016olt,Clough:2019jpm}, in violation of the no-hair theorem.
With a single event, \detector{} could constrain the dipole flux to a level of ${\sim}10^{-9}$, two orders of magnitude better than the best existing theory-agnostic constraints \cite{Barausse:2016eii, Kramer:2006nb}.
Thanks to its extreme sensitivity, \detector{} could also hope to directly detect or tightly constrain scalar or vector polarization modes in the GW strain \cite{Eardley:1974nw,Chatziioannou:2012rf}. In the case of electric charge, GW phase measurements could constrain the binary charge-to-mass ratio to $\lesssim 10^{-5}\sqrt{4\pi \epsilon_0 G}$~\cite{Barausse:2016eii,Cardoso:2016olt}. Thus, \detector{} would be able to constrain the neutrality of matter to the level of 1 excess electron per $10^{23}$ neutrons. Probing the neutrality of matter at this level tests models of (milli)-charged dark matter, and also informs us on the global and local neutrality of matter, a long-held but poorly tested belief.

\paragraph{Studying multipolar structure using inspiral.} Detailed measurements of the GW phase could also be used to probe the multipole moments of BHs. These moments may not be as anticipated if the objects we observe are not the astrophysical BHs predicted by GR, either because they are some other exotic object or because corrections to GR are needed \cite{TheLIGOScientific:2016src,Cardoso:2019rvt}. Detecting subtle deviations in the multipolar structure of these objects requires high-precision GW measurements. LISA will be able to measure up to two multipoles with good accuracy for comparable mass binaries, and possibly higher multipoles for intermediate mass ratios \cite{Kastha:2018bcr, Kastha:2019brk}. With greater sensitivity, \detector{} will enable the study of the detailed multipolar structure of detected binaries, for a greater number of sources.

\paragraph{Tidal Love numbers and BH mimickers.}
The tidal Love numbers of a star describe how the star responds to tidal forces, and encode the internal structure of the star. The Love numbers are imprinted on the GWs emitted during inspiral, and measurements of these will discriminate BHs from exotic compact objects which otherwise mimic the properties of BHs~\cite{Cardoso:2019rvt}.
\detector{} will detect longer inspirals and provide stringent measurements of these parameters, which are expected to vanish for BHs.

\subsection{The ringdown of supermassive BHs: gravitational spectroscopy }
Following the slow inspiral, SMBHs rapidly merge into a final BH. GWs emitted during the post-merger (``ringdown'') phase provide information about  strong gravity, dynamical spacetimes, and are uniquely clean probes of the dynamics of the final BH. Using only the least-damped (``fundamental'') ringdown mode, \detector{} can measure the remnant's mass and spin with a fractional error of $\mathcal{O}(10^{-5})$ for equal-mass mergers of $10^{6}-10^{7} M_\odot$ BHs at redshift $z=1$. Even greater precision could be achieved with overtones. 

\paragraph{Merging BHs.} Explorations of the plunge and merger phase of BHs in modified theories remain in their infancy. By the time \detector{} is launched, accurate numerical models in a variety of modified theories are expected to be available, and will allow for detailed comparison to the measured GWs. Currently, the ringdown phase following merger is understood in detail only in GR, and is certain to explore the geometry of spacetime near the BH horizon.

\paragraph{Parameterized ringdown tests.} Emission during ringdown can be described as the superposition of a number of damped oscillation modes, the QNMs of the BH.
The spectrum of modes depends only on the geometry of the final BH, which makes QNMs a powerful tool to explore the nature of BHs and constrain modified theories of gravity. 
A crucial element of ringdown studies is the measurement of multiple frequency modes, in terms of angular harmonics (quadrupole modes, octupole modes, etc.) and their corresponding overtones.
Even with the incorporation of multiple overtones~\cite{Giesler:2019uxc,Isi:2019aib}, ground-based GW detectors will measure multiple modes for relatively few events, and with relatively low SNR~\cite{Baibhav:2018rfk}. LISA will observe the ringdown following SMBHs with such high SNR that a considerable number of modes can be independently distinguished.

Following these pioneering studies, \detector{} will allow for detailed BH spectroscopy---the precision measurement of the structure of BHs using QNMs. By breaking several degeneracies, detection of large number of higher harmonics will also significantly improve sky localization and estimation of luminosity distance. With the measurements of QNM frequencies, direct constraints can be placed on the structure of spacetime near the BHs. QNM measurements can place constraints on the effective potential describing these modes and on their possible couplings~\cite{Cardoso:2019mqo,McManus:2019ulj}. LISA ringdown measurements  will provide constraints on the shapes of this potential, but the improved sensitivity of \detector{} will provide a detailed measurement of its angular and radial structure by measuring a greater number of QNMs for each SMBH binary.
This will be crucial for detecting or ruling out deviations from GR that only impact higher-order modes.

\section{Conclusions}
%

BHs are fascinating objects, related to a number of conundrums in fundamental physics, astrophysics, or cosmology. Long-standing unresolved issues, such as the nature of dark matter or dark energy, or the information paradox associated with BHs, force us to scrutinize gravity to new levels of detail, searching for answers. 

We have made a science case for \detector{}, a follow-up of LISA feasible to be built and launched in the 2035-2050s period. \detector{} would allow us to ``listen in'' deeper in the mHz gravitational-wave sky, gathering many more signals with higher precision, which would represent a major leap in probing the nature of BHs. Unparalleled insights could come from rare occurrences captured by the larger reach of \detector{}, the patterns emerging from the large statistics it would gather, or through the precise dissection of individual events. 

The LISA mission will leave a heritage in terms of technology, instrumentation, developed theoretical understanding, the strong surrounding community, and, finally, also the first millihertz GW sky map. We hope we have presented a convincing case that the \detector{} mission provides a unique chance to fully capitalize on this heritage and to foster and reinforce the exciting era of GW science that has just begun.

\newpage
\section*{References}
\begingroup
\renewcommand{\section}[2]{}%
\begin{multicols}{2}
\footnotesize
\bibliographystyle{apj_short_prop}
\bibliography{References}
\end{multicols}
\endgroup
\end{document}